\documentclass[12pt]{article}
\usepackage{amsmath, amssymb, amsthm, graphicx, geometry}
\usepackage[authoryear]{natbib}
\usepackage[ruled, linesnumbered, vlined]{algorithm2e}
\usepackage{hyperref}

\title{Discrepancy Modeling with Intermediate Variables: A New Framework for\\ Robust Gaussian Process Calibration}
\author{Henry Shaowu Yuchi$^{*}$, Michael Grosskopf$^{*}$, \\
Aman Sharma$^{\dagger}$, Nicolas Schunck$^{\dagger}$, \\
Jared O'Neal$^{\ddagger}$, Matt Menickelly$^{\ddagger}$, \\
Stefan M. Wild$^{\S}$\\
\\
$^{*}$Los Alamos National Laboratory\\
$^{\dagger}$Lawrence Livermore National Laboratory\\
$^{\ddagger}$Argonne National Laboratory\\
$^{\S}$Lawrence Berkeley National Laboratory
}
\date{}

\begin{document}

\maketitle

\begin{abstract}
Gaussian processes are widely used for surrogate modeling in computer experiments, which often produce numerous intermediate variables that are not explicitly used in standard calibration frameworks. Calibration of imperfect models can be challenging without leveraging these variables, while fitting the emulator and the discrepancy models separately also poses identifiability issues.
In this work, we propose a robust Gaussian process calibration framework that leverages intermediate variables for discrepancy modeling. The framework integrates a structured intermediate variable selection process, a discretized scaled Gaussian stochastic process (S-GaSP) to constrain the discrepancy term, and a space-filling design strategy for selecting constraint points. This enables joint modeling of the emulator and discrepancy, improving predictive performance, providing principled uncertainty quantification, and alleviating identifiability risks. 
We demonstrate its efficacy on a nuclear physics application involving binding energies, where it outperforms baseline approaches. 
\end{abstract}

\noindent%
\vfill

\newpage

\section{Introduction}
\label{sec:intro}
Running physical experiments are often extremely expensive and slow, due to the need to build enormous physical infrastructure or recreate extreme operating conditions. Computer experiments are therefore carried out to imitate the physical mechanisms and systems at a lower financial and time expenditure.
Scientists and engineers typically use complex codes and computer simulations to investigate large-scale and complicated real world problems across various topics, including nuclear physics \citep{langanke1991computational,salcedo1988computer,matta2016nptool}, computational fluid dynamics \citep{moraveji2017computational,wendt2008computational,nowak2003computational}, and geological activity analysis \citep{emery2007simulation,lin2012role,wu2020application}, to name a few. 
In such cases, system outputs are obtained given the values of input variables via simulations. They enable researchers to study these problems more efficiently, and open up the potential to utilize statistical modeling to acquire a better understanding of the physical systems to aid their investigation.
Emulator modeling is a powerful tool in computer experiments which mimics the relationship between simulation inputs and outputs \citep{sacks1989designs}, thus establishing a way to carry out predictions without running simulations.

There have been multiple ways to approach the emulator modeling, including, more recently, neural network-based models \citep{myren2021comparison}.
However, among these models the most popular and powerful one remains the Gaussian process (GP), which is versatile in its use to tackle emulation problems providing inference capability on complex systems \citep{williams1995gaussian,williams2006gaussian}. Crucially, it also provides the uncertainty quantification (UQ), which is often conducted via constructing a Bayesian model using the GP. 
The GP has been demonstrated to be effective in a wide range of settings and applications, including emulating the model of a cardiac cell \citep{chang2015bayesian}, the probabilistic flow of electrical power \citep{xu2020probabilistic}, and spatial modeling on inland flooding \citep{donnelly2022gaussian}. 

However, an emulator model depending on inputs alone is seldom potent enough to accurately represent the true relationship between inputs and outputs for complex physical systems. 
This is sometimes called an imperfect model \citep{sung2024efficient}.
In many experiments, there are underlying parameters in the system that have an inherent influence on the output. These unknown parameters often need to be adjusted to optimize the 
model's predictive performance. They are known as calibration parameters, and the process of fine-tuning the model using these parameters is called model calibration \citep{sung2024review, kennedy2001bayesian, santner2003design,plumlee2017bayesian}.
Model calibration is typically used to optimize the emulator model's predictive performance so that the model can be fine-tuned to better mimic the physical system. 
Calibration has been widely applied in different scientific studies ranging from epidemiology modeling \citep{sung2024efficient}, structural study of a mechanical part \citep{gattiker2006combining}, and cosmology \citep{higdon2013computer}. It has been especially popular in nuclear physics problems as well \citep{higdon2015bayesian,pratola2016bayesian}. 

A common challenge arises when the underlying computer model is an imperfect representation of the real world. In such a case, the optimal solution of the calibration process, e.g., the one that maximizes the likelihood or minimizes a generalized least square function, will still have limited accuracy.
For example, in nuclear density functional theory (DFT), which is used to compute the properties of atomic nuclei, the most common computer model is the solution to the Hartree-Fock-Bogoliubov (HFB) equation \citep{schunck2019energy} and the calibration parameters are a small set of coupling constants that characterize how the nuclear energy depends on the density of nucleons. Because the solution to the HFB equation is only an approximate computer model of the nucleus, it will be impossible to break some (unknown) accuracy threshold in reproducing real-world data. However, one can leverage domain-specific information to build an explicit model of the discrepancy between the results of the calibrated, imperfect model, and experimental measurements. A common practice is to build the discrepancy term using the same input variables for the physical model. 
Such discrepancy modeling is common in nuclear theory \citep{utama2016nuclear,utama2017refining,utama2018validating,niu2018nuclear,neufcourt2018bayesian,neufcourt2019neutron}. In recent studies, the authors built a robust discrepancy model with good extrapolation properties by leveraging the multiple intermediate output variables produced in the simulation outputs to achieve improved predictive performance for the calibrated physical model \citep{schunck2020calibration,perez2022controlling}.

In particular, discrepancy models that depend only on the system inputs may not fully capture biases introduced during calibration of an imperfect physical model. These biases can arise not only from the choice of experimental data, the form of the loss function, or the structure of the emulator, but also from the intrinsic limitations of the underlying simulator, which may drive calibration toward nonphysical regions of the parameter space. A promising approach is therefore to \emph{jointly} calibrate the parameters of the emulator model and the discrepancy model while incorporating intermediate variables. By exploiting additional physically meaningful information, this strategy has the potential to improve both predictive accuracy and robustness.


There are two major challenges in doing so: 
(i) Simulations typically produce a considerable number of intermediate variables with physical relevance. However, there is not yet an established process to select the best intermediate variables for discrepancy modeling to avoid potential extrapolation issues during inference; (ii) Joint parameter calibration of the emulator model and the discrepancy model can lead to identifiability issues between them.


To address these two challenges, in this work we propose a new framework for robust Gaussian process calibration, where intermediate variables are utilized to construct the discrepancy model. The proposed framework first carries out a variable selection process on the intermediate variables, which mitigates the extrapolation risk in inference.
It then utilizes a scaled Gaussian stochastic process (S-GaSP) to model the discrepancy. It constrains the magnitude of the discrepancy term, addressing the issue of identifiability between the emulator and the discrepancy models. The S-GaSP model is further coupled with a set of carefully designed constraint points to optimize its performance using a space-filling design of experiments method. 
Subsequently, the framework proposes a full Bayesian approach where the emulation and the discrepancy are jointly fitted. Therefore, a better predictive performance can be achieved while uncertainty quantification for calibration parameters are also enabled.

This article is organized as follows: In Section~\ref{sec:prelim}, we review the Gaussian process (GP) and discuss efforts made on discrepancy modeling to support GP emulation. Then in Section~\ref{sec:method} we introduce our proposed framework on discrepancy modeling with intermediate variables. With the proposed framework introduced, we apply it to the aforementioned nuclear physics study in Section~\ref{sec:apply}. Using this application, we compare the predictive performance of the proposed framework against the current baseline calibration model, demonstrating its superior performance and strength in describing uncertainty. 

\section{Preliminaries}
\label{sec:prelim}
In this section, we provide an overview of the Gaussian process, including its basic formulations and components, and how it is applied in emulation modeling. We touch on how the principal component analysis can be applied in GP modeling to mitigate the dimensionality issue.
We also introduce the fundamentals of discrepancy modeling, including different approaches to approach this problem, along with several potential issues. 
\subsection{Gaussian Process}
\label{sec:GP}
Computer experiments aim to imitate the physical systems by building an emulator modeling the relationship between its inputs and outputs.
In other words, denoting the $m$ system input variables by $\mathbf{x}\in\mathbb{R}^m$ and the corresponding output by $y\in\mathbb{R}$, the emulator aims to obtain an estimation function of the true underlying relation between system inputs and outputs, which is denoted by $f$:
\begin{equation}
y\approx f(\mathbf{x}).
\label{eq:emulator}
\end{equation}
Gaussian process (GP) has been a powerful tool typically used in modeling the emulation in computer experiments \citep{williams1995gaussian,williams2006gaussian}.
It is a multivariate stochastic process defined on the input space $\mathbb{R}^m$ where any finite collection of these $n$ input variables in the process follow multivariate Gaussian distributions. It starts with defining a GP prior distribution on the functions of the input space, which is described as follows:
\begin{equation}
f(\mathbf{x}) \sim \mathcal{GP}\left( \mu(\mathbf{x}), k(\mathbf{x},\mathbf{x}')\right).
\end{equation}
Here $\mu(\mathbf{x})$ is the mean function of the GP process, and $k(\mathbf{x},\mathbf{x}')$ is the kernel function that structures the covariance between function outputs at two arbitrary locations $\mathbf{x}$ and $\mathbf{x}'$ in the input space. For the kernel function $k$, one of the most popular kernel functions, the squared exponential kernel, is formulated as follows:
\begin{equation}
k(\mathbf{x},\mathbf{x}') = \alpha \exp \left\{ -\sum_{i=1}^m \frac{(x_i-x_{i'})^2}{\rho_i^2} \right\},
\label{eq:kernel-sq}
\end{equation}
where $\alpha$ is the marginal deviation of the process, and $\rho_i$ are the correlation lengths in the $i$-th dimension, defining the shape of the correlation function. Here they are regarded as hyper-parameters of the GP model.
Other choices of the kernel function can be used as well, including the Mat\'ern kernel \citep{melkumyan2011multi}. 
More recently, there have been efforts looking into nonstationary kernels to allow more flexibility in fitting the GP model \citep{montagna2016computer, paciorek2003nonstationary, plagemann2008nonstationary}. 
The GP model enables efficient mean and variance estimates with closed-form expressions. See Appendix~\ref{app:gp-posterior} for details.
The GP emulator can be fit either by maximum log-likelihood or utilizing a full Bayesian approach, where the model parameters are estimated via Gibbs sampling obtaining samples using Markov Chain Monte Carlo \citep{gelfand1990illustration, neal1997monte}.

There are several known constraints that can affect the efficacy of GP models. 
The computational cost of obtaining the conditional posterior distributions is heavily impacted by the calculation involving covariance matrices, which brings about the issue of scalability. In other words, fitting a GP model could get incredibly slow when there are a large number of points ($n$ is large), or the number of variables in inputs gets high ($p$ is large).
There have been multiple efforts to alleviate these issues. When the number of samples $n$ gets large, one school of thoughts is to impose a sparsity assumption on the GP model such that an approximation can be found \citep{snelson2012variable, lawrence2002fast, snelson2007local}. 
On the other hand, when the number of input variables $p$ gets high, a common remedy is to reduce the dimension of the input. 
A latent variable model based on principal component analysis (PCA) was first introduced to transform the input to achieve lower dimensions \citep{lawrence2005probabilistic} with an extension on Bayesian training procedure in \citep{titsias2010bayesian}.

Alternatively, when the number of output variables $q$ is high, PCA can also be applied to reduce its dimensionality. This approach, often referred to as PCA-GP, projects the outputs onto a low-dimensional basis and models the resulting latent variables using Gaussian processes \citep{higdon2008computer,higdon2015bayesian,schunck2020calibration}.
This scheme speeds up the training and fitting of the GP model when there are many output variables involved in the computer experiment, which alleviates the scalability concerns. 
A detailed PCA-GP algorithm for high-dimensional outputs is provided in Appendix~\ref{app:pca-gp}.

\subsection{Discrepancy Modeling}
\label{sec:disc}
There are different ways to approach calibration for emulator models, and in this work we focus on the discrepancy, which is an additive term on top of the emulator. The discrepancy model aims to capture the difference between the emulator model and physical reality with an additional discrepancy term. Building on the emulator model in \eqref{eq:emulator}, the discrepancy approach models the system output by
\begin{equation}
y = f(\mathbf{x}, \boldsymbol{\theta}) + \delta(\mathbf{x}) + \epsilon, \quad \epsilon\sim\mathcal{N}(0, \sigma^2),
\label{eq:disc}
\end{equation}
where $\boldsymbol{\theta}$ denote the calibration parameters for the emulator term, and $\sigma^2$ denotes the variance of the random noise $\epsilon$. The additional discrepancy term dependent on the system inputs $\delta(\mathbf{x})$ is introduced to gap the unexplained errors by the emulator model. This term is often modeled using a Gaussian process as well. A popular discrepancy model has been proposed in \cite{kennedy2001bayesian}, where $\delta$ is modeled by a zero-mean Gaussian process with the square exponential correlation function similar to Eqn.~\eqref{eq:kernel-sq}.
Therefore, the overall model in Eqn.~\eqref{eq:disc} is still a Gaussian process, facilitating parameter estimation.

In practice, the emulator $f(\mathbf{x}, \boldsymbol{\theta})$ is often fit first, and subsequently the discrepancy term would be fitted using the difference between the observations and the emulator $y-f(\mathbf{x}, \boldsymbol{\theta})$. However, this raises three issues which may limit the insight we may extract from the data: (i) When the emulation and the discrepancy terms are fit separately, we risks potential identifiability issues between the two terms which can lead to bias in prediction; (ii) The ability to exploit the GP for a full Bayesian analysis and carry out uncertainty quantification for both discrepancy and emulator parameters is lost; (iii) The discrepancy depends on the inputs $\mathbf{x}$ only, limiting its flexibility.

There have been efforts on obtaining the model calibration parameters by minimizing the $l_2$ distance between emulator and the observation. This idea is first proposed in \cite{tuo2015efficient, tuo2016theoretical}, where the calibration parameters are estimated via $L_2$ distance projection. It is formulated as follows:
\begin{equation}
\boldsymbol{\theta}^* = \text{arg}\min_{\boldsymbol{\theta}} \| y-f(\mathbf{x},\boldsymbol{\theta}) \|_{L_2(\Omega)},
\label{eq:l2-min}
\end{equation}
where $\Omega$ denotes the input space. Utilizing $l_2$ distance minimization, we may control the magnitude of the discrepancy model in fitting, mitigating potential identifiability issues. This addresses the first issue, but the latter two remain unsolved. This partly motivates our proposed method.

\section{Discrepancy Model via Intermediate Variables}
\label{sec:method}
In the nuclear physics problem briefly outlined in Section~\ref{sec:intro}, the computer simulations generate the outputs of the emulator. These include the binding energies across different nuclei and a group of intermediate variables, which bear clear physical meanings and can be well interpreted. 
Therefore, we propose a new discrepancy model utilizing these intermediate variables to improve the accuracy and interpretability of the discrepancy model, such that the predictions are more accurate when compared to real observations in binding energies. The formulation of the proposed model is as follows:
\begin{equation}
y = f(\mathbf{x},\boldsymbol{\theta})+\delta(\boldsymbol{\nu}(\mathbf{x},\boldsymbol{\theta}), \boldsymbol{\phi})+\epsilon,\quad \epsilon\sim\mathcal{N}(0,\sigma^2).
\label{eq:disc_model}
\end{equation}
Here, the discrepancy term is denoted by $\delta(\boldsymbol{\nu}(\mathbf{x},\boldsymbol{\theta}), \boldsymbol{\phi})$ with intermediate variables selected for the discrepancy model denoted by $\boldsymbol{\nu}$ which are dependent on the inputs $\mathbf{x}$ and calibration parameters $\boldsymbol{\theta}$, and $\sigma^2$ denotes the variance of the random error $\epsilon$. The parameters of the discrepancy term are denoted by $\mathbf{\phi}$. 
Specifically, the emulator term $f(\mathbf{x},\boldsymbol{\theta})$ is modeled by a Gaussian process with squared exponential kernels, and the discrepancy term $\delta(\boldsymbol{\nu}(\mathbf{x},\boldsymbol{\theta}), \boldsymbol{\phi})$ is modeled by a discrete scaled Gaussian stochastic process. This enables uncertainty quantification on the intermediate variables on top of predictions.


\subsection{Intermediate Variable Selection}
\label{sec:var_select}
When fitting discrepancy models using Gaussian processes, it is often desirable to restrict the number of intermediate variables involved in order to avoid overfitting and improve interpretability. When a computer experiment produces a large number of intermediate variables, efficient selection of these variables becomes essential.
In this work, we approach intermediate variable selection under the heredity principle, originally developed in regression modeling and experimental design \citep{wu2011experiments}. In particular, we adopt the strong heredity rule proposed by \cite{chipman1996bayesian} and extend it to discrepancy modeling with intermediate variables.

\textbf{Heredity Principle for Discrepancy Model}\\
If a discrepancy model involving multiple intermediate variables provides improved predictive performance, then each of those variables should also be effective when used individually in the discrepancy model. Formally, if
\[
\delta(\nu_1, \nu_2, \boldsymbol{\phi})
\]
is an effective discrepancy model, then both
\[
\delta(\nu_1, \boldsymbol{\phi}) \quad \text{and} \quad \delta(\nu_2, \boldsymbol{\phi})
\]
should also be effective.
Here, an ``effective discrepancy model'' refers to one that leads to a statistically significant reduction in prediction error compared to a model without the discrepancy term. That is, the residual
\[
y - f(\mathbf{x}, \boldsymbol{\theta}) - \delta(\boldsymbol{\nu}(\mathbf{x}, \boldsymbol{\theta}), \boldsymbol{\phi})
\]
is substantially smaller than
\[
y - f(\mathbf{x}, \boldsymbol{\theta}).
\]
Based on this principle, we propose the following sequential variable selection procedure:
\begin{enumerate}
\item Construct discrepancy models using each intermediate variable individually. Suppose there are $S$ intermediate variables $\{\nu_s\}_{s=1}^S$. For each $s$, fit
\[
y = f(\mathbf{x}, \boldsymbol{\theta}) + \delta^{(s)}(\nu_s, \boldsymbol{\phi}) + \epsilon.
\]
Select the variables that yield effective discrepancy models and collect their indices into the set $\Omega_1 \subset \{1,2,\dots,S\}$.

\item Form pairwise combinations of variables in $\Omega_1$ and construct discrepancy models of the form
\[
y = f(\mathbf{x}, \boldsymbol{\theta}) + \delta^{(u)}(\nu_{u_1}, \nu_{u_2}, \boldsymbol{\phi}) + \epsilon,
\]
where $(u_1, u_2)$ are selected from $\Omega_1$. Retain the effective combinations and store them in a set $\Omega_2$. This procedure can be extended to higher-order combinations, yielding sets $\Omega_3$, and so on.
\item Terminate the procedure when adding additional variables does not lead to a meaningful improvement in predictive performance. The final discrepancy model as in Eqn.~\eqref{eq:disc_model} is then constructed using the selected combination of intermediate variables.
\end{enumerate}
In practice, multiple candidate combinations may be identified. Since intermediate variables often carry physical meaning, domain knowledge can be incorporated to guide the final selection.

\subsection{Scaled Gaussian Stochastic Process}
\label{sec:s-gasp}
When the discrepancy model is defined on intermediate variables rather than the original system inputs, additional challenges arise. For emulators to perform well, it is generally desirable that the distribution of training data closely matches that of the prediction points. This is because most emulation methods, including Gaussian processes, perform significantly better under interpolation than extrapolation. Since the discrepancy model is constructed in the space of intermediate variables, extrapolation in this space can degrade predictive performance. In particular, the emulator may become overfitted in regions where training data are dense while performing poorly in regions with sparse coverage, thereby exacerbating identifiability issues between the emulator and the discrepancy model.

To address these challenges, we incorporate a constrained discrepancy formulation based on the discretized scaled Gaussian stochastic process (S-GaSP) model proposed by \cite{gu2018scaled}. For notational simplicity, we write $\delta(\boldsymbol{\nu}(\mathbf{x}))$ as $\delta(\boldsymbol{\nu})$. The S-GaSP model can be expressed as
\begin{equation}
    \delta(\boldsymbol{\nu}) = \mu(\boldsymbol{\nu}) + \left\{ \eta(\boldsymbol{\nu}) \,\middle|\, \sum_{i=1}^{N_C} \eta(\nu_i^C)^2 \Delta \nu = Z \right\},
    \label{eq:sgasp-simple}
\end{equation}
where $\eta$ is a zero-mean Gaussian stochastic process, $\{\nu_i^C\}_{i=1}^{N_C} \subset \mathcal{V}$ are a set of $N_C$ constraint points in the domain of the intermediate variables, and $Z$ is a constraint variable. This constraint effectively regularizes the discrepancy by penalizing large $l_2$ deviations from the emulator. The term $\Delta \nu = \text{Vol}(\mathcal{V})/N_C$ represents the discretization weight over the domain $\mathcal{V}$. The mean function $\mu(\boldsymbol{\nu})$ is typically specified as a linear function of $\boldsymbol{\nu}$.

The constraint variable $Z$ regulates the overall magnitude of the stochastic component $\eta$, preventing the discrepancy term from becoming excessively large. This helps alleviate identifiability issues between the emulator and the discrepancy model by discouraging the discrepancy from absorbing variations that should be attributed to the calibrated physical model. More details of S-GaSP, including how the prior distribution for $Z$ is configured, are included in Appendix~\ref{app:scaled-gp}.

The choice of constraint points plays an important role in the performance of the S-GaSP model. While these points can be selected to coincide with the observed intermediate variables for computational convenience, the resulting distribution is often non-uniform. Therefore, careful selection of constraint points is necessary to ensure adequate coverage of the domain and to avoid extrapolation in regions with sparse data.

\subsection{Design of Experiments for Constraint Points}
\label{sec:doe}
The constraint points in the discrete S-GaSP model are introduced to approximate an integral term via discretization (see Eqn.~\eqref{eq:sgasp-simple}). This implies that an effective set of constraint points should be approximately uniformly distributed over the space of intermediate variables $\mathcal{V}$. In practice, however, intermediate variables obtained from computer experiments are often unevenly distributed, making direct reuse of these points suboptimal. To achieve accurate discretization and avoid extrapolation, the constraint points need to provide adequate coverage of $\mathcal{V}$ and are therefore ideally selected independently of the observed intermediate variables.

The selection of constraint points can be naturally framed within the context of design of experiments. In particular, we seek a space-filling design over $\mathcal{V}$. A variety of methods have been proposed to construct such designs \citep{joseph2016space}, including Latin hypercube designs and maximum projection designs. These approaches aim to generate samples that are approximately uniformly distributed over the design space, thereby improving the accuracy of the discretized S-GaSP model. In this work, we adopt a Latin hypercube design to construct the constraint set.
With the discrete S-GaSP discrepancy model and an appropriately chosen set of constraint points, the full predictive distribution of the model, incorporating both the emulator and discrepancy, can be derived for inference and uncertainty quantification.

In practice, several additional considerations arise. First, the size of the constraint set $N_C$ directly impacts computational cost, as shown in Eqn.~\eqref{eq:var_disc} in Appendix~\ref{app:scaled-gp}. Large values of $N_C$ increase the cost of covariance matrix computations and can significantly slow down inference. Therefore, $N_C$ should be chosen carefully based on the dimension of the design space and available computational resources. 
Second, incorporating constraint points into the Bayesian inference procedure introduces additional computational cost, particularly within MCMC sampling. A practical compromise is to pre-estimate quantities involving the constraint points and treat them as fixed during sampling, trading off some accuracy for computational efficiency.
Finally, the choice of the regularization parameter $\lambda$ requires careful tuning. This parameter controls the influence of the constraint term and effectively regulates the magnitude of the discrepancy $\delta(\boldsymbol{\nu}(\mathbf{x}))$. In practice, $\lambda$ can be selected using cross-validation or other model selection criteria.

\subsection{Inference}
\label{sec:infer}
Given the emulator and the proposed discrepancy model based on intermediate variables using the scaled Gaussian stochastic process (S-GaSP), we consider the full model for inference as defined in Eqn.~\eqref{eq:disc_model}.
The emulator is modeled as a Gaussian process, while the discrepancy term is modeled using an S-GaSP. These two components are assumed to be independent.

We now derive the predictive distribution of the model.
Let $\mathbf{x}^*$ denote a new input location at which prediction is desired, and let $\boldsymbol{\nu}^*$ denote the corresponding intermediate variables. Assuming for the moment that $\boldsymbol{\nu}^*$ is known, the conditional posterior predictive distribution can be expressed as
\begin{equation}
    y(\mathbf{x}^*) \mid \mathbf{y} \sim \mathcal{N}(\hat{\mu}(\mathbf{x}^*), \sigma_\eta^2 c^* + \sigma^2),
\label{eq:final}
\end{equation}
where
\begin{align}
\begin{split}
    \hat{\mu}(\mathbf{x}^*) &= f(\mathbf{x}^*) + \mu(\boldsymbol{\nu}^*) + \mathbf{r}(\boldsymbol{\nu}^*)^T \tilde{\mathbf{R}}^{-1} \big( \mathbf{y} - \mathbf{f}(\mathbf{x}) - \mu(\boldsymbol{\nu}) \big),\\
    c^* &= c(\mathbf{x}^*, \mathbf{x}^*) - \mathbf{r}(\mathbf{x}^*)^T \tilde{\mathbf{R}}^{-1} \mathbf{r}(\mathbf{x}^*).
\end{split}
\label{eq:final_detail}
\end{align}
Here, $\tilde{\mathbf{R}} = \mathbf{R} + \sigma^2 \mathbf{I} / \sigma_\eta^2$, and $\mathbf{r}(\mathbf{x}^*)$ denotes the correlation vector between the new point and observed data, adjusted by the constraint structure of the S-GaSP model, leading to $\mathbf{r}(\mathbf{x}^*)=\mathbf{r}^\eta(\mathbf{x}^*)-(\mathbf{r}^C)^T(\tilde{\mathbf{R}}^C)^{-1}\mathbf{r}^C$. Here, $r(\cdot)$ and $\tilde{R}$ denote correlation structures induced by the S-GaSP model; full expressions are provided in Appendix~\ref{app:scaled-gp}.
Since the predictive distribution is also a Gaussian, point estimates and uncertainty quantification can be obtained directly. In particular, credible intervals for predictions can be constructed from the posterior distribution.

In practice, the intermediate variables at new input locations, $\boldsymbol{\nu}^*$, are typically unknown and must be estimated. To address this, we introduce an additional Gaussian process model for the mapping $\boldsymbol{\nu}(\mathbf{x})$:
\begin{equation}
    \boldsymbol{\nu} \sim \mathcal{GP}(\mu_{\boldsymbol{\nu}}(\boldsymbol{\nu}), k_{\boldsymbol{\nu}}(\boldsymbol{\nu}, \boldsymbol{\nu}')),
\end{equation}
where the mean and the kernel functions are parametrized. In the inference procedure, this model is trained simultaneously with the rest of the framework. As a result, for a new input $\mathbf{x}^*$, we can obtain an estimate $\hat{\boldsymbol{\nu}}(\mathbf{x}^*)$, which is then used in the discrepancy model for prediction. With all components specified, the full Bayesian inference algorithm is summarized in Appendix~\ref{app:algo_full}. Predictions at arbitrary inputs can then be obtained by evaluating the fitted emulator and discrepancy model.

\section{Application: Calibrate Nuclear Energy Functionals}
\label{sec:apply}
We now apply the proposed framework to a nuclear physics problem and revisit the calibration of nuclear energy functionals \citep{schunck2020calibration} as introduced in Section~\ref{sec:intro}. We conduct a simulation study based on the Skyrme energy density functional (EDF) to assess the predictive performance of the proposed discrepancy model. The primary objective of this application is to build a calibrated emulator model with a discrepancy term to predict the observed binding energies for different nuclei. This is achieved by obtaining (i) an emulation model for the relation between simulation inputs and simulated binding energies as outputs across different nuclei; and (ii) a discrepancy model to account for the difference between the emulator predictions and the observed binding energies. Here we denote observed physical outputs by $y_{\text{obs}}$, corresponding to the generic response variable $y$ in the model formulation in the previous section.

The calibration parameters, a small set ($p=12$) of coupling constants denoted by $\boldsymbol{\theta}\in\mathbb{R}^{12}$, are traditionally obtained by minimizing a $\chi^2$ cost function
containing the scaled difference between experimental measurements and computed values of selected properties of a small set of atomic nuclei; details about the form of the $\chi^2$ function, the experimental dataset, and the computer model can be found in \cite{kortelainen2012nuclear, schunck2020calibration}.
The simulation input contains the proton and neutron numbers of a set of $n=75$ nuclei, which are used to identify the isotopes. Thus the inputs can be denoted by $\mathbf{x}\in\mathbb{R}^{75\times 2}$.
The simulation output contains the (simulated) binding energy $E_j$ over the same atomic nuclei $j=1,2,\dots,n$. We denote it by $\mathbf{y}_{\text{sim}}=\left[E_1, E_2,\dots, E_n\right]\in\mathbb{R}^{75}$. The emulator models the output by $f(\mathbf{x},\boldsymbol{\theta})$. 
Each simulation produces a range of intermediate physical variables which include the deformation parameters characterizing the shape of the nuclei, the RMS-radius of the nuclei, pairing energies, pairing gaps, and kinetic energies. See Table~\ref{tbl:intermediate} in Appendix~\ref{app:int-list} for a comprehensive list of intermediate variables.

In this study, a set of $Q=500$ simulations are generated by varying the EDF parameters $\boldsymbol{\theta}$ to produce simulated binding energies for the $n=75$ nuclei to be used for constructing the emulator. 
The calibration parameters corresponding to the $Q$ simulations are denoted by $\boldsymbol{\theta}_k$, where $k=1,\dots,500$.
For each simulation, we have the same inputs $\mathbf{x}\in\mathbb{R}^{75\times 2}$ and the corresponding outputs $\mathbf{y}_{k}\in\mathbb{R}^{75}$. 
The simulated dataset is then employed to fit the emulator model, while the observed dataset is used to learn a discrepancy model formulated as a function of intermediate variables, enabling extrapolation beyond the 75 observed nuclei.

To evaluate predictive performance of the proposed framework, we consider a larger validation dataset of $n^* = 885$ nuclei for which observed binding energies are available. These experimental binding energies are obtained from the atomic mass evaluation (AME) \citep{huang2021ame,wang2021ame}. Predictions for this validation set are obtained by propagating posterior samples of the calibration parameters $\boldsymbol{\theta}$ through the calibrated emulator and discrepancy model. Model performance is then assessed by comparing these predicted binding energies to the observed values in the validation set.


\subsection{Emulation without Discrepancy}

\begin{figure}[!t]
\centering
    \includegraphics[width=.99\linewidth]{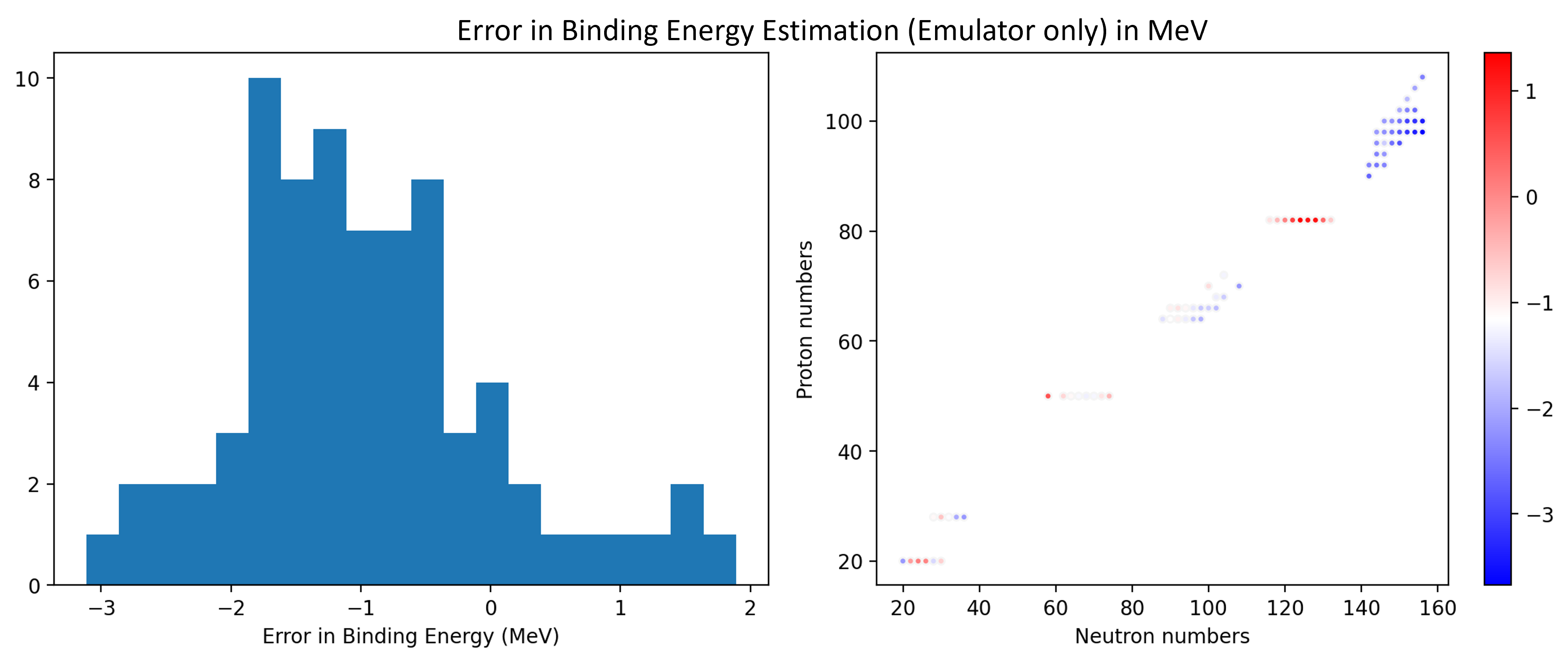}
    \caption{Estimation errors in binding energy of training set only using the emulator.}
    \label{fig:no_disc}
\end{figure}

We start with fitting the GP-based emulator model alone using the simulation set, utilizing the emulator model alone:
\begin{equation}
    \label{eq:nodisc}
    y_{\text{obs}} = f(\mathbf{x},\boldsymbol{\theta}) + \epsilon,
\end{equation}
where $\epsilon\sim\mathcal{N}(0,\sigma_\epsilon^2)$ denotes the inherent uncertainty, and $\boldsymbol{\theta}$ denotes the calibration parameters, which are coupling constants of the EDF \citep{kortelainen2012nuclear}. From the fitted emulator, we plot the errors between the observed and the emulated binding energy for the training set, expressed as $y_{obs, n}-f(\mathbf{x}_n,\boldsymbol{\theta})$ for $n\in\{1,\dots,75\}$, with respect to the proton (Z) and neutron (N) numbers of the nuclei in the training set, which is illustrated in Figure~\ref{fig:no_disc}. The left plot illustrates the histogram of the errors, and the right one illustrates the errors with respect to neutron and proton numbers. From the histogram of errors on the left, we observe that there is a clear bias in the estimate, which is confirmed by the calculated mean square error which is 2.261 MeV. 
From the plot on the right, we observe that there is consistent and patterned bias, especially when the proton numbers become large. This indicates that the emulator model alone cannot accurately capture all the structures and patterns in the experimentally observed data. Therefore, the further step of the discrepancy term becomes necessary to reduce the bias to improve the quality of observed binding energy predictions. 
Additionally, the distribution of neutron (N) and proton (Z) numbers in the training set indicates that building the discrepancy terms defined on $\mathbf{x}$ will lead to extrapolation which is detrimental to predictive accuracy. It is necessary to utilize the intermediate variables instead.


\subsection{Discrepancy Modeling with Physical Variables}

Since the intermediate variables we obtain in addition to binding energy outputs from the simulations bear clear physical meanings, we attempt to utilize them in building the discrepancy model for calibration. In the application, there are $\approx 70$ intermediate variables obtained in simulations. However, only 26 of them possess interpretable physical implications as listed in Appendix~\ref{app:int-list}, which become the focus of our study.

We conduct the variable selection process to identify the best combinations of physical variables to be used in the model calibration. We start with only using one individual physical variable to construct the model including the discrepancy term as follows:
\begin{equation}
    y_{\text{obs}} = f(\mathbf{x},\boldsymbol{\theta})+\delta(\boldsymbol{\nu}(\mathbf{x}, \boldsymbol{\theta}), \boldsymbol{\phi}) + \epsilon,
\end{equation}
where the discrepancy is modeled using the discretized scaled Gaussian stochastic process. In this application, we assume a zero-mean discrepancy, i.e., $\mu(\nu)=0$, for simplicity.

\begin{figure}[!t]
\centering
    \includegraphics[width=.99\linewidth]{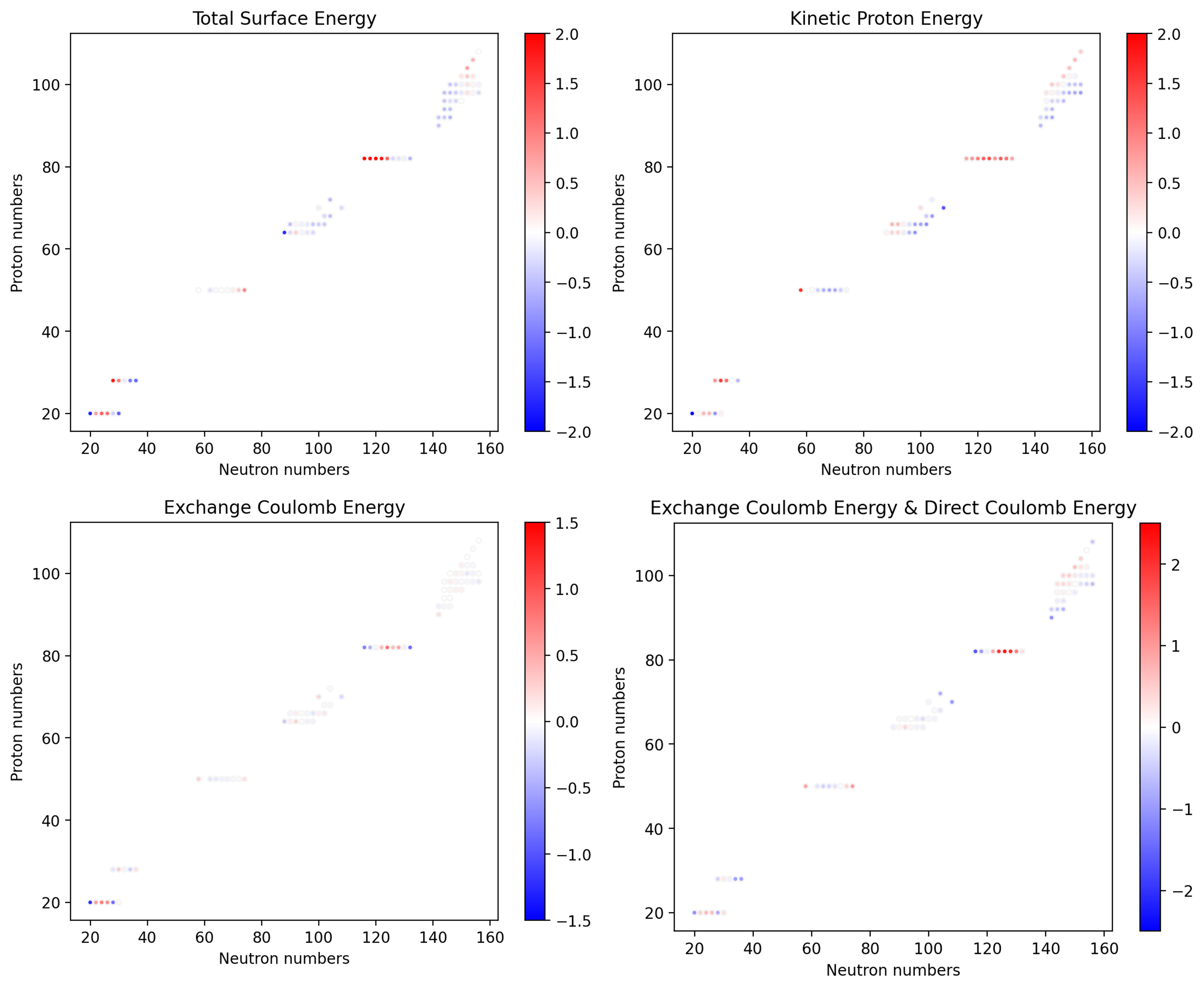}
    \caption{Estimation errors in binding energy using emulator and discrepancy.}
    \label{fig:calibrate_1d}
\end{figure}

Each of the 26 intermediate physical variables is tested to evaluate the bias in estimating the observed binding energy for the training set of 75 nuclei. 
Three of them are plotted in Figure~\ref{fig:calibrate_1d} to illustrate how the variable selection procedure is carried out, where the title indicates the physical variable being used to construct the discrepancy. 
The binding energy estimation errors (in MeV) after applying discrepancy models with physical intermediate variables are plotted with respect to the N and Z numbers.


From the plots, we observe that the prediction errors in the case of exchange Coulomb energy are significantly smaller than in the emulator-only case. This suggests that calibrating the model using discrepancy built on exchange Coulomb energy is effective. On the other hand, the fitting residuals become larger for both kinetic proton energy and total surface energy. This indicates that both are ineffective choices to be selected for the discrepancy model. Furthermore, in all three cases, we can still see some structural bias when plotted against neutron and proton numbers, suggesting there are still patterns not captured by the discrepancy term. To summarize, across all physical intermediate variables, we obtain through the screening procedure the following five variables that have proved effective: exchange Coulomb energy, direct Coulomb energy, kinetic proton energy, total surface energy, and neutron deformation $\beta_2$.

Utilizing the heredity principle for variable selection, we then fit updated discrepancy models where there are now two physical variables contained in $\boldsymbol{\nu}$, both selected from the aforementioned list of effective individual variables. From them we have identified a best combination which includes exchange Coulomb energy and direct Coulomb energy, and exhibits smaller residuals without a clear remaining pattern. The resulting estimation errors are shown in the bottom-right plot of Figure~\ref{fig:calibrate_1d}.

Among all two-variable combinations, this pair of exchange Coulomb energy and direct Coulomb energy shows the most significant improvement in estimation errors. 
Furthermore, we check selecting three physical variables for the discrepancy model from the selected set. However, it no longer produces further improved results for residuals anymore, so we settle with using two physical variables for calibration in this application.

\subsection{Model Evaluation}
Now that the intermediate variables for the discrepancy model $\boldsymbol{\nu}$ are determined, we perform Bayesian inference for the proposed framework and evaluate its predictive performance on observed binding energies.  
We compare the proposed approach against baseline calibration models using the same inputs $\mathbf{x}$, with predictions obtained by propagating posterior samples of the calibration parameters $\boldsymbol{\theta}$.  
The discrepancy term is modeled using a discrete scaled Gaussian stochastic process, dependent on the selected intermediate variables $\boldsymbol{\nu} \equiv \boldsymbol{\nu}(\mathbf{x}, \boldsymbol{\theta})$.

Due to the discretization carried out in scaled Gaussian stochastic process (S-GaSP), it is necessary to determine the set of constraint points. Given two intermediate physical variables have been selected for the discrepancy model, we constructed a 64-point Latin hypercube design in the 2D space for $\boldsymbol{\nu}$, which is space-filling.

\begin{figure}[!hb]
\centering
    \includegraphics[width=\linewidth]{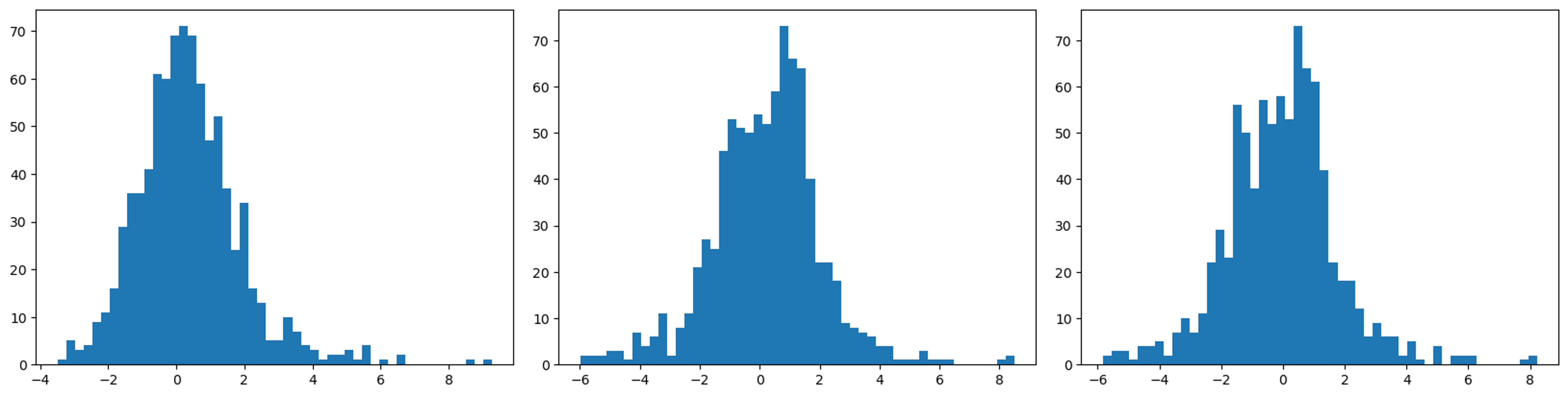}
    \caption{Histogram plots of prediction errors in binding energy.}
    \label{fig:hist_final}
\end{figure}

With the constraint point set obtained, we move forward to evaluate the predictive performance of the proposed framework on the validation set. Numerically, we evaluate the performance by calculating the Root Mean Square Error (RMSE) between the predicted binding energy and the observed values as shown below:
\begin{equation}
    \text{RMSE} = \sqrt{\frac{\sum_{i=1}^{n'} (y_{\text{obs},i}-\hat{y}_i)^2 }{n'}},
\label{eq:rmse}
\end{equation}
where $y_{\text{obs},i}$ denotes the observed binding energy for nucleus $i$ in the validation set, and $\hat{y}_i$ denotes the corresponding predicted binding energy.
There are $n'=885$ nuclei in the validation set. 
For comparison, we utilize two baseline methods which reflect common practices in calibration and discrepancy modeling. The first baseline uses the emulator model without the discrepancy term, as in Eqn.~\eqref{eq:nodisc}. The second baseline fits the emulator model first and then fits the discrepancy term on the residual. It is formulated as follows:
\begin{equation}
y_{\text{obs}}=f(\mathbf{x},\boldsymbol{\theta})+\zeta (\boldsymbol{\nu})+\epsilon,
\end{equation}
where we fit the emulator model $f(\mathbf{x},\boldsymbol{\theta})$ first and then use the residual to fit the discrepancy term $\zeta(\boldsymbol{\nu})$. This procedure is distinctly different from the proposed method in fitting $\delta$. We use the two baselines and the proposed model to finally predict the observed binding energy in the validation set.
The histogram plots of the errors between the estimated and the true observed binding energies are plotted in Figure~\ref{fig:hist_final}. Here the discrepancy model uses two selected physical intermediate variables: exchange Coulomb energy and direct Coulomb energy. The left plot corresponds to the proposed framework where both emulation and discrepancy terms are fitted simultaneously. The plot in the center corresponds to the first baseline where only the emulation term is fitted (without discrepancy). The right plot corresponds to the baseline case where the emulation term is fitted first, then the discrepancy is fitted on residuals subsequently.

Comparing the predictions to the true values, we obtain that the RMSE for the proposed method in predicting the observed binding energies is 1.59 MeV, while the RMSE for the non-discrepancy baseline method is 1.82 MeV. The proposed method achieves an approximately 14\% reduction in prediction error, which demonstrates the efficacy of bringing in the discrepancy term. On the other hand, the RMSE for the second baseline method is 1.80 MeV, which is 13\% higher than the proposed method. This supports the use of intermediate variables to construct the discrepancy term is beneficial. 
At the same time, the results show the advantages of fitting the emulation and the discrepancy simultaneously rather than sequentially, where we now acquire the ability to mitigate potential identifiability issues and to obtain uncertainty quantification of model parameters.


Additionally, the error histogram plots in Fig.~\ref{fig:hist_final} show that the proposed method has prediction errors that are more symmetrically centered around zero, whereas the baselines exhibit slightly biased and less regular error distributions. This indicates lower systematic bias compared to the two baseline methods and suggests that the proposed framework captures residual structure more effectively, leading to improved predictive accuracy. The results further suggest that the proposed method alleviates the identifiability issue between the emulation and discrepancy terms in calibration.

Finally, we examine the posterior distributions of the calibration parameters to assess uncertainty quantification under the proposed framework. The posterior samples are obtained jointly with the discrepancy model, enabling full Bayesian inference on the calibration parameters. The distributions are provided in Appendix~\ref{app:posterior}. We observe the posterior distributions of the calibration parameters under the proposed framework exhibit tighter and more stable lobes than those obtained under emulator-only calibration. This suggests that the regularized discrepancy model improves parameter identifiability by reducing confounding between the emulator and discrepancy components.

\section{Discussions and Conclusions}
In this work, we study surrogate modeling for computer experiments that generate intermediate variables during simulation. Tackling the imperfect emulator model, and to address the lack of systematic methods for incorporating such variables into discrepancy modeling, as well as the identifiability challenges in joint calibration, we propose a robust Gaussian process framework that explicitly leverages intermediate variables.

The proposed framework combines a structured intermediate variable selection procedure, a discretized scaled Gaussian stochastic process (S-GaSP) for discrepancy modeling, and a space-filling design strategy for selecting constraint points for the discrepancy. This enables joint modeling of the emulator and discrepancy, leading to improved predictive performance and principled uncertainty quantification. The efficacy of the approach over baseline methods is demonstrated through its application to a nuclear physics problem, where it shows superior predictive performance over baseline approaches.

A core advantage of the proposed robust GP framework is that it breaks a lack of identifiability between the discrepancy model and emulation parameters. The assumption inherent in this model is that the physical model predictions should be consistent with the data and the discrepancy capture the remaining variability. This is a common assumption in cases where the model is fit to the data and discrepancy modeled post-hoc \citep{perez2022controlling}. However, this is a data-constrained assumption, not an assumption based on physical constraints. For instance, in their ``simple machine'' example in \citep{brynjarsdottir2014learning}, the authors outline an example not uncommon with physical models, where the missing term in their model form results in a systematic over-prediction of the data by the model at the 'true' parameters. An analogous physical problem is a model which neglects drag in a ballistics simulation. In this work, the robust GP model presented would prefer parameters that allowed the model to go through the data, rather than the ``correct'' parameters where the model over-predicts the data and the discrepancy corrects in one direction. The authors conclude that use of subject matter expertise is critical in properly regularizing the discrepancy in these cases, and incorporation of that expertise with the robust GP model is the subject of future work.

Going forward, we plan to focus on streamlining the variable selection process for intermediate variables by looking into the horseshoe prior configuration. We would also like to explore the physical implications using the results from uncertainty quantification on the nuclear physics applications.

\newpage
\bibliographystyle{plain}
\bibliography{references_v2}

\newpage
\begin{center}

{\large\bf SUPPLEMENTARY MATERIAL}

\end{center}
\setcounter{section}{0}
\section{PCA-Gaussian Process Algorithm}
\label{app:pca-gp}
{\small
\begin{algorithm}[H]
\LinesNumbered
\caption{PCA-Weighted Gaussian Process (\texttt{PCA-GP})}
\label{alg:PCA-GP}
\KwData{Training inputs: $\mathbf{X}\in\mathbb{R}^{n\times m}$.\\
\quad\quad\quad \hspace{0.000005in} Training outputs: $\mathbf{Y}\in\mathbb{R}^{n\times \ell}$.}
\textbf{Parameters:} No. of input points: $n$; Input dimension: $m$;\\
\quad\quad\quad\quad\quad\quad \hspace{0.025in} Output dimension: $\ell$; No. of PCA weights selected: $D$.\\
\KwIn{Testing input: $\tilde{\mathbf{x}}\in\mathbb{R}^m$.}
Standardize the outputs by column: $\mathbf{Y}_{\text{std}} \leftarrow (\mathbf{Y}-\bar{\mathbf{Y}})/\sigma_{\mathbf{Y}}$;\\
Carry out PCA: $\mathbf{Y}_{\text{std}}\approx \mathbf{W}\mathbf{K}^T$;\\
Normalize principal weights by column:
$\mathbf{W}_{\text{std}}\leftarrow\mathbf{W}/\sigma_{\mathbf{W}}$;\\
Rescale principal vectors by column: $\mathbf{K}_{\text{scale}}\leftarrow\mathbf{K} \, \sigma_{\mathbf{W}}$;\\
\textbf{Train one GP per PCA weight}:\\
\For{$i=1:D$}{
    Fit a scalar GP model $w_i(\mathbf{x})\sim\mathcal{GP}(m_i(\mathbf{x}), k_i(\mathbf{x},\mathbf{x}'))$,\\
    where $\mathbf{W}_{\text{std}[:,i]}=\left( w_i(\mathbf{x}_1),\dots, w_i(\mathbf{x}_n) \right)^T$;
}
Obtain predicted mean from GP: $\bar{w}(\tilde{\mathbf{x}})=(w_1(\tilde{\mathbf{x}}), \dots, w_D(\tilde{\mathbf{x}}))^T$;\\
Obtain predicted variance: $\text{Var}[\bar{w}(\tilde{\mathbf{x}})] = \text{diag}(\text{Var}[w_1(\tilde{\mathbf{x}})],\dots, \text{Var}[w_D(\tilde{\mathbf{x}})])$;\\
Reconstruct predicted output: $\hat{y}(\tilde{\mathbf{x}})\leftarrow \mathbf{K}_{\text{scale}}\bar{w}(\tilde{\mathbf{x}})\odot\sigma_{\mathbf{Y}} + \bar{\mathbf{Y}}$;\\
Reconstruct predicted variance: $\Sigma(\tilde{\mathbf{x}}) \leftarrow \text{diag}(\sigma_{\mathbf{Y}})\mathbf{K}_{\text{scale}}\cdot\text{Var}[\bar{w}(\tilde{\mathbf{x}})] \cdot\mathbf{K}_{\text{scale}}^T\cdot\text{diag}(\sigma_{\mathbf{Y}})$;\\
\KwOut{Predicted testing output: $\hat{\mathbf{y}}(\tilde{\mathbf{x}})$; Predicted testing variance: $\Sigma(\tilde{\mathbf{x}})$;\\
\quad\quad\quad\quad\hspace{0.02in} PCA weights and vectors $\mathbf{W}_{\text{std}}, \mathbf{K}_{\text{scale}}$; \\
\quad\quad\quad\quad\hspace{0.02in} output mean and standard deviation $\bar{\mathbf{Y}}, \sigma_{\mathbf{Y}}$;\\
\quad\quad\quad\quad\hspace{0.02in} Fitted GP models $w_i(\mathbf{x})$ for $i=1,2,\dots,D$.}
\end{algorithm}
}
\newpage
\section{Robust GP Calibration with Discrepancy}
\label{app:algo_full}
{\small
\begin{algorithm}[H]
\caption{Robust GP Calibration with Discrepancy}
\label{alg:disc-GP}
\KwData{Training inputs: $\mathbf{X}\in\mathbb{R}^{n\times m}$; Intermediate variables: $\boldsymbol{\nu}\in\mathbb{R}^{n\times s}$; Testing input: $\tilde{\mathbf{x}}\in\mathbb{R}^m$;\\
\quad\quad\quad\hspace{0.000002in} Training simulation outputs: $\mathbf{Y}\in\mathbb{R}^{n\times \ell}$; Training observed outputs: $\mathbf{y}_{\text{obs}}\in\mathbb{R}^{\ell}$;}
\textbf{Parameters:} No. of input points: $n$; Input dimension: $m$;\\
\quad\quad\quad\quad\quad\quad \hspace{0.025in} Output dimension: $\ell$; No. of PCA weights selected: $D$;\\
\quad\quad\quad\quad\quad\quad \hspace{0.025in} No. of intermediate variables: $s$; No. of calibration parameters: $p$.\\
Obtain $\mathbf{W}_{\text{std}}, \mathbf{K}_{\text{scale}}, \bar{\mathbf{Y}}, \boldsymbol{\sigma}_{Y}$, and $w_i(\mathbf{x}), i=1,2,\dots, D$ from \texttt{PCA-GP}$(\mathbf{X}, \mathbf{Y})$;\\
Obtain $\mathbf{W}_{\text{std}}^{\text{int}}, \mathbf{K}_{\text{scale}}^{\text{int}}, \bar{\mathcal{N}}^{\text{int}}, \boldsymbol{\sigma}_{\mathcal{N}}^{\text{int}}$, and $w_i^{\text{int}}(\mathbf{x}), i=1,2,\dots, D$ from \texttt{PCA-GP}$(\mathbf{X}, \boldsymbol{\nu})$;\\
Initialize calibration parameters $\boldsymbol{\theta}\sim\mathbf{U}[0,1]^p$;\\
Initialize discrepancy parameters $\alpha_{\delta}\sim\text{Gamma}(0.5, 0.5)$ and $\sigma\sim\text{Gamma}(5, 5)$;\\
\For{$i =1:s$}{
    Initialize discrepancy parameter $\rho_{i}\sim \text{InvGamma}(3,1)$;
}
\For{$t= 1:T$}{
    Compute $f(\boldsymbol{\theta}^{(t-1)}) \leftarrow \mathbf{K}_{\text{scale}}\bar{\mathbf{w}}(\boldsymbol{\theta}^{(t-1)}) \odot \boldsymbol{\sigma}_{Y} + \bar{\mathbf{Y}}$;\\
    Compute $\Sigma_f(\boldsymbol{\theta}^{(t-1)}) \leftarrow \operatorname{diag}(\boldsymbol{\sigma}_{Y})\mathbf{K}_{\text{scale}}\cdot \operatorname{Var}[\bar{\mathbf{w}}(\boldsymbol{\theta}^{(t-1)})]\cdot \mathbf{K}_{\text{scale}}^T \cdot \operatorname{diag}(\boldsymbol{\sigma}_{Y})$;\\
    Compute $\boldsymbol{\nu}(\boldsymbol{\theta}^{(t-1)}) \leftarrow \mathbf{K}_{\text{scale}}^{\text{int}}\bar{\mathbf{w}}^{\text{int}}(\boldsymbol{\theta}^{(t-1)}) \odot \boldsymbol{\sigma}_{\mathcal{N}}^{\text{int}} + \bar{\mathcal{N}}^{\text{int}}$;\\
    Compute $\mu_{\delta}(\boldsymbol{\theta}^{(t-1)})$ and $\Sigma_{\delta}(\boldsymbol{\theta}^{(t-1)})$ from $\boldsymbol{\nu}(\boldsymbol{\theta}^{(t-1)})$ via Eqn.~\eqref{eq:mean_disc} and \eqref{eq:var_disc};\\
    Compute $\Sigma_{\sigma}^{(t-1)} \leftarrow (\sigma^{(t-1)})^2 \mathbf{I}_{\ell}$;\\
    Form the posterior distribution: $p(\boldsymbol{\theta}, \boldsymbol{\rho}, \alpha_{\delta}, \sigma \mid \mathbf{y}_{\text{obs}})
    \propto
    \mathcal{N}\!\left(\mathbf{y}_{\text{obs}}; f(\boldsymbol{\theta}) + \mu_{\delta}, \Sigma_f + \Sigma_{\delta} + \Sigma_{\sigma}\right)
    p(\boldsymbol{\theta}) p(\boldsymbol{\rho}) p(\alpha_{\delta}) p(\sigma);$\\
    Sample $\boldsymbol{\theta}^{(t)} \sim p(\boldsymbol{\theta}\mid \boldsymbol{\rho}^{(t-1)}, \alpha_{\delta}^{(t-1)}, \sigma^{(t-1)}, \mathbf{y}_{\text{obs}})$;\\
    Sample $\boldsymbol{\rho}^{(t)} \sim p(\boldsymbol{\rho}\mid \boldsymbol{\theta}^{(t)}, \alpha_{\delta}^{(t-1)}, \sigma^{(t-1)}, \mathbf{y}_{\text{obs}})$;\\
    Sample $\alpha_{\delta}^{(t)} \sim p(\alpha_{\delta}\mid \boldsymbol{\theta}^{(t)}, \boldsymbol{\rho}^{(t)}, \sigma^{(t-1)}, \mathbf{y}_{\text{obs}})$;\\
    Sample $\sigma^{(t)} \sim p(\sigma\mid \boldsymbol{\theta}^{(t)}, \boldsymbol{\rho}^{(t)}, \alpha_{\delta}^{(t)}, \mathbf{y}_{\text{obs}})$;\\
    Sample posterior predictive draw $\tilde{\mathbf{y}}^{(t)} \sim p(\tilde{\mathbf{y}} \mid \tilde{\mathbf{x}}, \boldsymbol{\theta}^{(t)}, \boldsymbol{\rho}^{(t)}, \alpha_{\delta}^{(t)}, \sigma^{(t)})$ via Eqn.~\eqref{eq:final} and \eqref{eq:final_detail};\\
}
Calculate posterior predictive mean and variance from sampled draws $\tilde{\mathbf{y}}^{(t)}, t=1,2,\dots,T$.\\
\KwResult{Posterior predictive mean $\hat{\mathbf{y}}$, Posterior predictive variance $\text{Var}[\hat{\mathbf{y}}]$.}

\end{algorithm}
}
\newpage
Note: For notational simplicity, we suppress the dependence of $f$, $\boldsymbol{\nu}$, and $\delta$ on the input locations $\mathbf{x}$; unless otherwise specified, all quantities are evaluated at the observed input locations, while predictions are evaluated at the testing input $\tilde{\mathbf{x}}$.

\section{GP Posterior Predictive Distribution}
\label{app:gp-posterior}
Suppose there are $n$ observed simulations where both the inputs are denoted by $\mathbf{x}_1,\mathbf{x}_2,\dots,\mathbf{x}_N$ and the corresponding simulation outputs are denoted by \\
$\bar{\mathbf{f}}=[f(\mathbf{x}_1),f(\mathbf{x}_2).\dots.f(\mathbf{x}_n)]^T$.
The GP model can be utilized to conduct predictions on simulation outputs at other locations in the input space. Using a GP prior, the posterior distribution at an arbitrary set of $R$ new input locations denoted by $\bar{\mathbf{x}}^*=[\mathbf{x}^*_1; \mathbf{x}^*_2;\dots; \mathbf{x}^*_R]\in\mathbb{R}^{R\times m}$ can be formulated. Denoting the inputs of the observation set by $\bar{\mathbf{x}}=[\mathbf{x}_1;\mathbf{x}_2;\dots;\mathbf{x}_p]\in\mathbb{R}^{n\times m}$, we obtain the following:
\begin{align}
\begin{split}
f(\bar{\mathbf{x}}^*) | \bar{\mathbf{f}}&\sim\mathcal{N}\left(\tilde{\mu}(\bar{\mathbf{x}}^*), \tilde{\Sigma}(\bar{\mathbf{x}}^*,\bar{\mathbf{x}}^*)\right) \\
\text{where }\tilde{\mu} (\bar{\mathbf{x}}^*) &= \mu(\bar{\mathbf{x}}^*) + \Sigma(\bar{\mathbf{x}}^*,\bar{\mathbf{x}}) \Sigma(\bar{\mathbf{x}},\bar{\mathbf{x}})^{-1}(\bar{\mathbf{f}} - \mu(\bar{\mathbf{x}}))\\
\tilde{\Sigma}(\bar{\mathbf{x}}^*,\bar{\mathbf{x}}^*) &= \Sigma(\bar{\mathbf{x}}^*,\bar{\mathbf{x}}^*)  -\Sigma(\bar{\mathbf{x}}^*,\bar{\mathbf{x}}) \Sigma(\bar{\mathbf{x}},\bar{\mathbf{x}})^{-1} \Sigma(\bar{\mathbf{x}}^*,\bar{\mathbf{x}})^T,
\end{split}
\end{align}
where $\Sigma(\bar{\mathbf{x}}^*, \bar{\mathbf{x}})\in\mathbb{R}^{R\times n}$, $\Sigma(\bar{\mathbf{x}}^*, \bar{\mathbf{x}}^*)\in\mathbb{R}^{R\times R}$, and $\Sigma(\bar{\mathbf{x}}, \bar{\mathbf{x}})\in\mathbb{R}^{n\times n}$ denote the correlation matrices.
The posterior distribution enables us to obtain both point estimates in closed-form analytical expression on the predicted outputs $f(\bar{\mathbf{x}}^*)$ given a number of observed simulations, and at the same time provide uncertainty quantification as well.

\section{Scaled Gaussian Stochastic Process}
\label{app:scaled-gp}
The scaled Gaussian stochastic process (S-GaSP) is formulated as follows:
\begin{align}
\begin{split}
    \delta_z(\boldsymbol{\nu}(\mathbf{x})) &= \left\{ \eta(\boldsymbol{\nu}) | \int_{\xi\in \mathcal{V}} \eta(\xi) ^2 d\xi=Z \right\},\\
    \eta(\cdot) &\sim \text{GaSP} (0, \sigma_\eta^2c^\eta(\cdot,\cdot)),\\
    Z&\sim p_{\delta_z}(\cdot).
\end{split}
\label{eq:SGaSP}
\end{align}
Here the term $\delta_z(\boldsymbol{\nu})$ denotes the scaled Gaussian stochastic process. It is assumed that $\eta$ follows a zero-mean Gaussian stochastic process (GaSP), where the probability density of samples $\{\boldsymbol{\nu}_1,\boldsymbol{\nu}_2,\dots, \boldsymbol{\nu}_D\}$ is a multivariate normal distribution. It can be expressed as below:
\begin{equation}
    \left( \eta(\boldsymbol{\nu}_1),\eta(\boldsymbol{\nu}_2),\dots, \eta(\boldsymbol{\nu}_D) \right)^T | \mathbf{R}^\eta \sim \mathcal{N} (0, \sigma_\eta^2\mathbf{R}^\eta),
\end{equation}
where $\mathbf{R}^\eta$ denotes the covariance matrix usually constructed from power exponential kernel functions as in Eqn.~\eqref{eq:kernel-sq}. It has been shown that the conditional distribution $p_\eta(Z|\boldsymbol{\psi})$ given the GaSP parameters $\boldsymbol{\psi}$ is the same as an infinite weighted sum of non-central chi-squared distributions \cite{gu2018scaled}, which can be approximated by discretization. Subsequently, the distribution of $Z$ can be chosen as follows:
\begin{equation}
p_{\delta_z}(Z=z|\boldsymbol{\psi}) 
= 
\frac{f_Z(Z=z|\boldsymbol{\psi})p_\eta(Z=z|\boldsymbol{\psi})}{\int_0^\infty f_Z(Z=t|\boldsymbol{\psi})p_\eta(Z=t|\boldsymbol{\psi})dt},
\end{equation}
where $f_Z$ is the conditional probability for $Z$. The measure of the random variable $Z=\int_{\xi\in \mathcal{V}} \eta(\xi) ^2 d\xi$ is critical for the S-GaSP model, which is also the $L_2$ loss between the physical truth and the emulator term. 

This shows that the distribution of the S-GaSP variable $\delta(\boldsymbol{\nu})$ is a scaled distribution of the GaSP variable $\eta$. The S-GaSP model aims to provide flexibility in modeling the system while at the same time keeps the $L_2$ loss between the physical reality and the emulator model adequately low. This effectively restricts the magnitude of the discrepancy term. It fits both the emulator model and the discrepancy model simultaneously, which mitigates the identifiability issue between the emulator and the discrepancy terms when fitted separately.

Applying the S-GaSP model to the discrepancy term, we utilize its discretized version in this work, which substitutes the integral in the condition in \eqref{eq:SGaSP} with a summation. The discretized S-GaSP model is therefore expanded from Eqn.~\eqref{eq:sgasp-simple} as follows:
\begin{align}
\begin{split}
    \delta(\boldsymbol{\nu}) &= \mu(\boldsymbol{\nu})+\left\{ \eta(\boldsymbol{\nu}) | \sum_{i=1}^{N_C} \eta(\nu_i^C) ^2 \Delta \nu=Z \right\},\\
    \eta(\cdot) &\sim \text{GaSP} (0, \sigma_\eta^2c^\eta(\cdot,\cdot)),\\
    Z&\sim p_{\delta}(\cdot).
\end{split}
\label{eq:disc-SGaSP}
\end{align}
Here, the integral is discretized with $N_C$ distinct points $\nu_i^C\in\mathcal{V}$ for $i=1,2,\dots,N_C$, with $\Delta\nu=\text{Vol}(\mathcal{V})/N_C$ and $\text{Vol}(\mathcal{V})$ being the volume of the domain for intermediate variables $\boldsymbol{\nu}$. An additional mean term $\mu(\boldsymbol{\nu})$ can be added here which is usually a linear combination of $\boldsymbol{\nu}$.
We call this set of $N_C$ points the constraint points.
The marginal distribution of $\delta(\boldsymbol{\nu})$ follows a multivariate normal distribution:
\begin{equation}
    \delta(\boldsymbol{\nu})|\boldsymbol{\psi} \sim \mathcal{N}(\mu(\boldsymbol{\nu}),\sigma_\eta^2\, \mathbf{R}),
    \label{eq:mean_disc}
\end{equation}
where
\begin{equation}
    \mathbf{R} = \mathbf{R}^\eta - (\mathbf{r}^C)^T\left( \mathbf{R}^C + \frac{N_C}{\lambda} \mathbf{I}_{N_C}  \right)^{-1} \mathbf{r}^C.
    \label{eq:var_disc}
\end{equation}
Here $\mathbf{R}^C$ is the correlation matrix for the constraint points, and $\mathbf{r}^C$ is the cross-correlation matrix between the constraint points and the points in the original dataset, both using the square exponential kernel. 

\section{List of intermediate physical variables}
\label{app:int-list}
There are 26 intermediate physical variables that are generated during UNEDF1 binding energy simulations in Section~\ref{sec:apply} which are candidates for $\boldsymbol{\nu}$, and can be used to construct the discrepancy model. These variables are listed in Table~\ref{tbl:intermediate} as follows.

\begin{table}[htbp]
\centering
\caption{Intermediate physical variables generated from UNEDF1 binding energy simulations.}
\begin{tabular}{cc}
\multicolumn{2}{c}{Table of Relevant Intermediate Variables} \\ \hline\hline
Neutron Fermi energy          & Proton Fermi energy          \\
Neutron pairing gap           & Proton pairing gap           \\
Neutron pairing energy        & Proton pairing energy        \\
Neutron rms-radius            & Proton rms-radius            \\
Neutron deformation $\beta_2$ & Proton deformation $\beta_2$ \\
Neutron quadrupole moment     & Proton quadrupole moment     \\
Neutron hexadecapole moment   & Proton hexadecapole moment   \\
Neutron multipole moment $q_{60}$  & Proton multipole moment $q_{60}$ \\
Neutron multipole moment $q_{80}$  & Proton multipole moment $q_{80}$ \\
Neutron kinetic energy        & Proton kinetic energy        \\
Volume energy ($\rho\tau$ contribution) & Volume energy ($\rho\rho$ contribution)\\
Surface energy                & Spin-orbit energy            \\
Direct Coulomb energy         & Exchange Coulomb energy     
\end{tabular}
\label{tbl:intermediate}
\end{table}

\section{Posterior Distributions of Calibration Parameters}
\label{app:posterior}
In this section, we look into the posterior distributions of the calibration parameters $\boldsymbol{\theta}$ obtained from the proposed joint calibration framework. These posterior samples are obtained via the full Bayesian inference procedure described in Section~\ref{sec:infer}, and the posterior distribution of $\boldsymbol{\theta}$ are plotted in Fig.~\ref{fig:UQ} below. The left plot shows the distributions obtained from the emulator-only calibration model without discrepancy, and the right plot shows the distributions for the joint calibration case proposed.
\begin{figure}[!hb]
\centering
    \includegraphics[width=\linewidth]{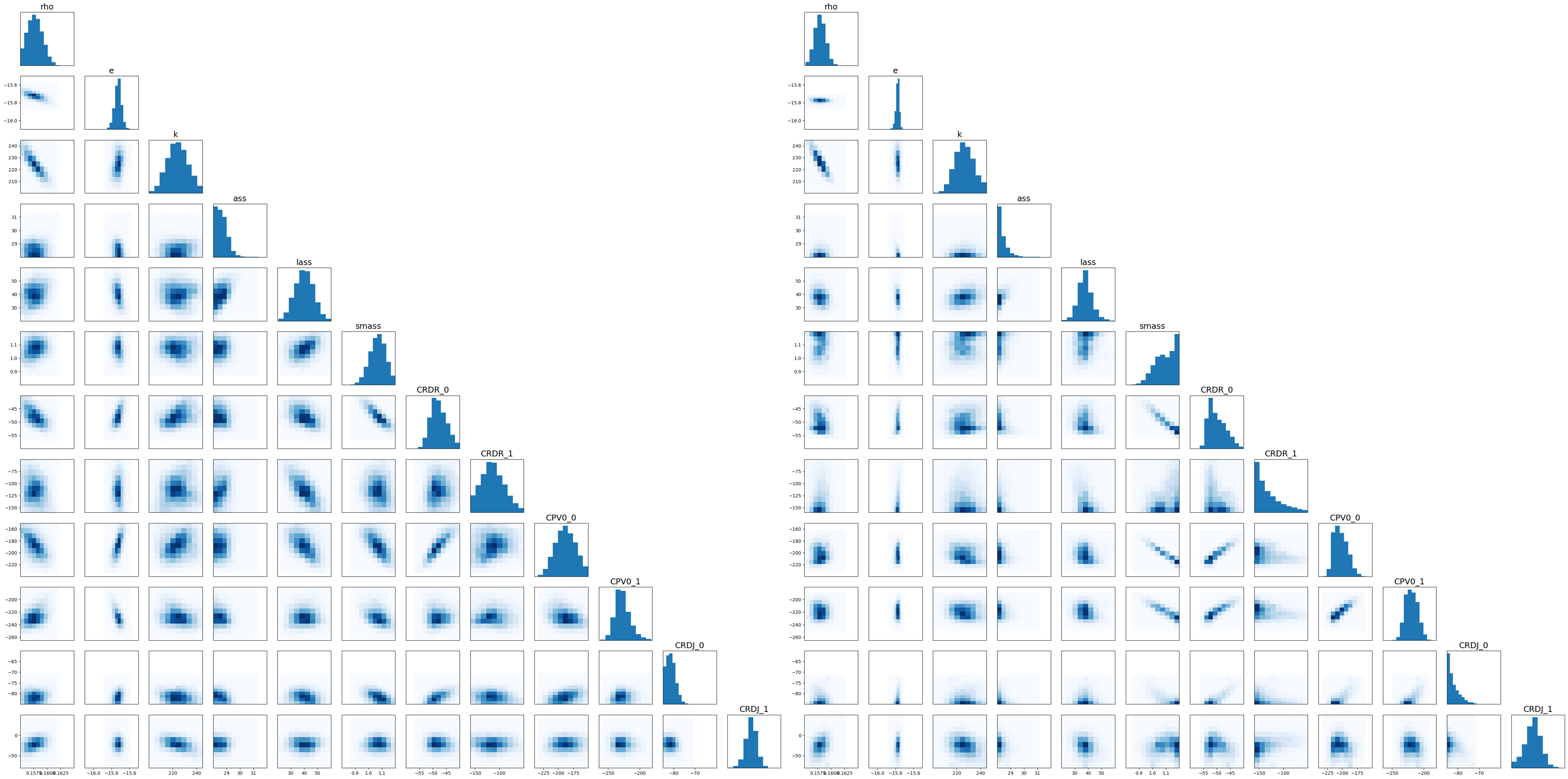}
    \caption{Distribution of calibration parameters.}
    \label{fig:UQ}
\end{figure}

From the plots, we observe that the proposed framework yields more concentrated and stable posterior distributions for several calibration parameters compared to the emulator-only approach. 
This behavior is consistent with the regularizing effect of the S-GaSP discrepancy model, which reduces confounding between the emulator and discrepancy components.
On the other hand, the posterior distributions remain sufficiently spread to reflect uncertainty in the calibration process, indicating that the model does not become overly restrictive. The results support the effectiveness of the proposed joint calibration framework in providing meaningful uncertainty quantification for the calibration parameters.
\end{document}